\documentclass[aps,twocolumn,showpacs]{revtex4-1}
\usepackage{amsmath,amsfonts,amssymb,wrapfig}
\DeclareFontFamily{U}{rsfs}{}         
\DeclareFontShape{U}{rsfs}{m}{n}{<5> rsfs5 <6><7> rsfs7          %
  <8><9><10><10.95><12><14.4><17.28><20.74><24.88> rsfs10}{}     %
\DeclareMathAlphabet{\mathfs}{U}{rsfs}{m}{n}                     %
\newcommand{\mfs}[1]{\mathfs {#1}}                               %
\def\beq{\begin{eqnarray}}
\def\eeq{\end{eqnarray}}

\def\bg{{\bf g}}
\def\nn{\nonumber\\}

\def\del{\partial}
\def\grad{\nabla}

\def\lie{\mfs L}

\begin{document}
\title{Entropy from near-horizon geometries of Killing horizons}
\author{Olaf Dreyer}\email{olaf.dreyer@gmail.com}
\affiliation{Dipartimento di Fisica, Universit\`a di Roma ``La Sapienza" and Sez. Roma1 INFN,\\
Piazzale Aldo Moro 2, 00185 Roma, Italy}
\author{Amit Ghosh}\email{amit.ghosh@saha.ac.in}
\author{Avirup Ghosh}\email{avirup.ghosh@saha.ac.in}
\affiliation{Saha Institute of Nuclear Physics, 1/AF Bidhan Nagar, Kolkata 700064, India.}

\begin{abstract}
We derive black hole entropy based on the near-horizon symmetries of black hole space-times. To derive these symmetries we make use of an $(R,T)$-plane close to a Killing horizon. We identify a set of vector fields that preserves this plane and forms a Witt algebra. The corresponding algebra of Hamiltonians is shown to have a non-trivial central extension. Using the Cardy formula and the central charge we obtain the Bekenstein-Hawking entropy.  
\end{abstract}

\maketitle

\section{Introduction}
After Bekenstein and Hawking \cite{BK,HW} had shown that black holes possess an entropy the question arose whether a statistical mechanical explanation could be given for it. More precisely, what are the micro-states of a black hole that are responsible for its entropy? A possible clue came from the fact that the black hole temperature has its origin in quantum mechanics (it contains a factor of $\hbar$ and in the naive classical limit $\hbar\to 0$ the temperature vanishes). This indicates that the micro-states in question are also quantum mechanical in origin. However, because a quantum theory of gravity remains elusive, so is the exact nature of these micro-states. Recently some progress has been made both in string theory \cite{strings} and loop quantum gravity \cite{lqg,GhoshPerez} to identify effective micro-states for black holes close to the horizon). However, in the absence of a satisfactory quantum theory of black holes it is not surprising that alternative methods (which do not depend on the details of a specific model of quantum gravity) have also been suggested.  

In this paper we follow one such alternative approach, the symmetry based approach to black hole entropy \cite{Strominger1,Strominger2,Carlip1}. The main idea behind this approach is to identify a symmetry group that we expect to be present in any quantum theory of black holes. The states of this quantum theory, which are possibly the black hole micro-states in question, would furnish a representation for this symmetry group. The representation is expected to be characterized by the relevant black hole parameters such as horizon area, charges, angular momentum etc. The black hole entropy would then be given by the dimension of such a characteristic representation (for example, the number of quantum states for a given area, angular momentum, charge etc.). From the outset, the aim is to calculate just the number of micro-states, knowing that from the symmetry group alone we will not be able to identify the micro-states. This method has been very popular and successful in obtaining entropy for a wide class of black holes \cite{Cvitan:2002rh,Majhi:2011ws,Majhi:2012tf,Zhang:2012fq}.

This approach to black hole entropy goes back to the classic work of Brown and Henneaux \cite{BH}. They found two copies of the Virasoro algebra as the asymptotic symmetry algebra of $2+1$ dimensional space-times that are asymptotically anti-de Sitter. It was Strominger who realized that the dimension of the representation of this Virasoro algebra gives the right Bekenstein-Hawking entropy for the BTZ black hole \cite{Strominger1}. However, the symmetries in \cite{BH,Strominger1} are associated with asymptotic infinity.  It was thus unclear how relevant this observation is to the question of black hole entropy since the reasoning applies to any space-time, black hole or not,  that is asymptotically anti-de Sitter.  This issue was cleverly addressed by Carlip, who found that a similar set of symmetries exists near the horizon of a black hole \cite{Carlip1,Carlip2} (see also \cite{OAW}). 



The problem that arises when one looks for symmetries near the horizon is that the usual notion of symmetries has to be expanded. The set of symmetries of the metric represented by the Killing vectors is not enough (this is also applicable for the analysis of symmetries at asymptotic infinity) \cite{Bondi:1962px, Sachs:1961zz,Penrose:1962ij,Newman:1966ub,Ashtekar:1978zz,Ashtekar:1984zz,Ashtekar:1999jx}. We therefore use an extended notion of symmetries that allows for a larger set of vector fields. We will do this by demanding that the diffeomorphisms generated by these vectors fields leave the near-horizon geometry invariant (in the appropriate sense). We will use this criterion along with some other mild assumptions to fix the vector fields. Once the near-horizon symmetries are completely specified the question remains what is the 
ensemble of space-times that remains invariant under the diffeomorphisms generated by these vector fields? We then find an ensemble such that each space-time of the ensemble falls off appropriately to a given geometry close to the horizon. Then we find that the algebra of these vector fields is isomorphic to the Witt algebra Diff $(S^{1})$.

The identification of these vector fields is simplified by the introduction of a two-dimensional plane that exists for all non-extremal Killing horizons in arbitrary dimensions. In the simple case of a Schwarzschild black hole this plane is the plane spanned by the Schwarzschild coordinates $r$ and $t$. Because of this we will call this plane the $(R,T)$-plane. The geometry of this plane contains all the essential features that mark a space-time with a black hole. The desired vector fields are expected to keep the geometry of this plane invariant. We then extend these vector fields to the entire space-time. These extended vector fields form the same Witt algebra.

After we have identified the algebra of vector fields we look for the corresponding Hamiltonian generators in the phase space and calculate their Poisson brackets using the appropriate symplectic structure \cite{silva}. We find that this algebra possess a non-vanishing central extension. Using this central charge and the Cardy formula \cite{Cardy} we calculate the entropy which is in agreement with the Bekenstein-Hawking entropy for black hole. 

Our paper consists of three parts. In the first part we introduce and study the $(R,T)$-plane for a non-extremal Killing horizon in any dimension. In the second part, we demonstrate its universal near-horizon geometry and find the algebra of these vector fields which is isomorphic to Diff $(S^1)$. In the third part, we represent this algebra by the corresponding Hamiltonian generators in the phase space and find that the algebra is centrally extended. This gives us the entropy of the black hole.

\section{The $(R,T)$-plane}

Consider a $(n+1)$-dimensional space-time $(\mfs M,\bg)$ that admits a Killing horizon $\Delta$. The Killing vector field $T^a$ is null on $\Delta$ and time-like in a neighbourhood $\mfs N\subset\mfs M$ of $\Delta$ ($\mfs N$ belongs to one side of $\Delta$ only where $T^a$ is time-like). For example, if $\mfs M$ is a Schwarzschild black hole then $\mfs N$ is an open subset of the space-time region from asymptotic infinity up to the event horizon. The space-time region inside the black hole will not concern us here. If the norm of the Killing vector field $T^a$ is denoted by $T$ then we define a one-form $\rho$ by
\beq \rho = -\frac{1}{2}dT^2.\label{rho}\eeq
Let $R^a$ to be the vector field associated with $\rho$, such that for an arbitrary vector field $X$,
\beq \bg(R,X) = \rho(X).\label{rdef}\eeq
Since $T$ is Killing it satisfies the Killing equation
\beq \bg(X, \nabla_Y T) + \bg( Y, \nabla_X T) = 0,\label{killing}\eeq
for an arbitrary pair of vector fields $X$ and $Y$. Then (\ref{rho}), (\ref{rdef}) and (\ref{killing}) implies $R$ and $T$ are orthogonal:
\beq \bg(R,T) = 0.\label{orthogonal}\eeq
Proof: From (\ref{rho}) and (\ref{rdef}), $\bg(R,T)=-\bg(T,\grad_TT)$. Then in (\ref{killing}), put $X=Y=T$. $\Box$

Moreover, since $\rho$ is exact, it is also closed:
\beq d\rho = 0.\label{eqn:drhozero}\eeq
Using $d\rho(X,Y)=X[\rho(Y)]-Y[\rho(X)]-\rho([X,Y])$
along with (\ref{rdef}) and (\ref{eqn:drhozero}), it follows that $R$ satisfies an identity:
\beq \bg(X, \nabla_Y R) - \bg( Y, \nabla_X R) = 0.\label{eqn:negkilling}\eeq
Using (\ref{eqn:negkilling}), (\ref{killing}) and (\ref{orthogonal}), we find that $R$ and $T$ commute. Proof: For an arbitrary vector field $X$,
\begin{align} \bg( X, \nabla_T R) & =  \bg( T, \nabla_X R)
	 = - \bg( \nabla_X T, R) \nn
	& = \bg( X, \nabla_R T ). \nonumber\quad\Box
\end{align}
Thus,
\beq [R,T] = 0. \label{commute}\eeq
The vector fields $R$ and $T$ are thus integrable. We will call the two-dimensional plane spanned by $R$ and $T$ the $(R,T)$-plane. In what follows we shall be concerned with the properties of this $(R,T)$-plane only.

The $(R,T)$-plane squeezes into a line on the horizon $\Delta$. Proof: We can see this by deriving another expression for $R$. For an arbitrary vector field $X$ the Killing equation (\ref{killing}) and metric compatibility give
\begin{align} \bg(R,X)&=-\frac{1}{2}\nabla_XT^2
 	=-\bg(T,\nabla_XT)
	=\bg(\nabla_TT,X)\nonumber
\end{align}
Thus,
\beq R=\nabla_TT.\eeq
On the horizon $\nabla_TT\triangleq\kappa T$ (here $\triangleq$ means that the equality on $\Delta$ only), where $\kappa$ is the surface gravity of the horizon. So the vector field $R$ becomes parallel to $T$ on $\Delta$: $R\triangleq\kappa T$. $\Box$

For $\kappa\neq 0$ (non-extremal horizons) $R$ is proportional to $T$ on $\Delta$ and for $\kappa=0$ (extremal horizons) $R$ vanishes on $\Delta$. As $T$ is time-like in $\mfs N$ and $R$ is orthogonal to $T$ and non-vanishing, it must be space-like in $\mfs N$.

\section{The geometry of the $(R,T)$-plane}

Since $R$ and $T$ commute, we can introduce coordinates $r$ and $t$ in the $(R,T)$-plane such that
\beq R = \frac{\partial}{\partial r}\ \ \ \mbox{and}\ \ \ T = \frac{\partial}{\partial t}. \eeq
Moreover, since $\rho(T)=0$, $T^2$ is a function of $r$ only. Let us define a smooth function $\omega(r)$ by
\beq T^2 = -\kappa^2\omega^2(r). \eeq
Since $T$ is null on the horizon, it follows that $\omega$ vanishes on the horizon, $\omega\triangleq 0$. We will use $\omega$ as a coordinate to measure distances from the horizon.

The metric or line element $ds^2_{RT}$ on the $(R,T)$-plane is of the form (since $R$ and $T$ are orthogonal)
\beq ds^2_{RT} = T^2 dt^2 + R^2 dr^2. \label{rtplane}\eeq
The norm $R^2$ of $R$ is given by
\begin{align} R^2 = \bg(R,R) = -\frac{1}{2}\nabla_RT^2
	= \kappa^2\omega\omega^\prime.
\end{align}
The metric $ds^2_{RT}$ is thus given by
\beq ds^2_{RT} = -\kappa^2\omega^2 dt^2 + \kappa^2\frac{\omega}{\omega'} d\omega^2. \eeq

We now want to determine the ratio $\omega/\omega'$ in the vicinity of the horizon, i.e., where $|T^2|\ll 1$. It is basically the ratio, $T^2/R^2=-\omega/\omega'$. To calculate the ratio, we assume that $R=\nabla_TT$ is differentiable in $T^2$ and can be expanded in the form
\beq R=\nabla_TT=\kappa T+T^2X+o(T^4)Y,\label{expans}\eeq
where $X,Y$ are vector fields. The ratio of $R^2/T^2$ is then given by
\beq R^2/T^2=\kappa^2+2\kappa\bg(T,X)+o(T^2).\label{ratio}\eeq
However, from the expansion (\ref{expans}) and (\ref{orthogonal}),
\beq 0=\bg(T,R)=T^2(\kappa+\bg(T,X))+o(T^4);\eeq
which shows that $\bg(T,X)=-\kappa+o(T^2)$. As a result,
\beq R^2=-\kappa^2T^2+o(T^4)\label{fratio}.\eeq
Equating (\ref{fratio}) with $R^2=\omega\omega'$ and for $\omega^2\ll 1$ we get
\beq \frac{\omega}{\omega'}=\frac{1}{\kappa^2}+o(\omega^2).\eeq
Thus, neglecting terms of order $o(\omega^2)$,
\beq ds^2_{RT}=-\kappa^2\omega^2dt^2+d\omega^2.\label{mihp}\eeq
The above analysis shows that this $(R,T)$-plane exists in the near horizon region of a Killing horizon having a non-vanishing surface gravity in arbitrary dimensions.

The plane (\ref{mihp}) can be Wick-rotated ($\tau=it$) to form a Euclidean plane where $\kappa\tau$ plays the role of an angle whose periodicity has to be $2\pi$ in order to avoid a conical deficit. This gives the $\tau$-period to be $\beta=2\pi/\kappa$ -- the correct one that reproduces the Hawking temperature.

Clearly, this plane captures the essence of the near-horizon geometry of a stationary black hole. We will consider the Euclidean geometry in the following discussions 
\beq ds^2=\kappa^2\omega^2d\tau^2+d\omega^2\label{rte}\eeq
in which $T=\del/\del\tau$ and $R=\del/\del r=\omega'\del_{\omega}$.

\section{Near Horizon Symmetries}

The near-horizon symmetry is defined so that the geometry of the $(R,T)$-plane is preserved. More precisely, a vector field $\xi=A\del_\tau+B\del_\omega$ in the $(R,T)$-plane where $A,B$ are smooth functions of $\tau,\omega$ will be said to preserve the $(R,T)$-plane if the following conditions are satisfied: {\tt (i)} $(\lie_\xi{\bf g})(T,T)=0$, {\tt (ii)} ${\bf g}(\lie_\xi R,T)=0$. These two conditions give the following class of vector fields:
\beq \xi_A=A\frac{\partial}{\partial\tau}-\omega\del_\tau A\frac{\partial}{\partial\omega},\quad A=A(\tau).\label{xi}\eeq
The first condition preserves the notion of proximity to the horizon and the second one the orthogonality of $R$ and $T$. It then follows for the above class of vector fields that both $(\lie_\xi\bg)(R,T)$ and $\bg(R,\lie_\xi T)$ are $o(\omega^2)$. Moreover, $[\lie_\xi R,T]=[R,\lie_\xi T]=o(\omega^3)$. The Killing vector field $T$ must be a member of this class of symmetry generating vector fields $\xi$. So it makes no sense of preserving it -- all what we can expect is that $\lie_\xi T$ remains in this class and we find that $\lie_{\xi_A}T=-\xi_{T(A)}$. We also find that all the fall-offs are at least $o(\omega^2)$, the same order to which the metric (\ref{rte}) is written down. Thus, the geometry of the plane is preserved. The vector field $R=\omega^\prime\partial_\omega$ is a normal derivative to the horizon, the second condition also means that it remains so under the diffeomorphisms generated by $\xi_A$. 

Starting from a given background, the diffeomorphisms generated by the vector fields $\xi_A$ gives an ensemble of space-times that share the same near-horizon geometry of the $(R,T)$-plane in the precise sense defined above. Since the $\kappa$-defining equation $\grad_TT\triangleq\kappa T$ is invariant under diffeomorphisms that are tangential to the horizon and $\kappa$ is a constant on the horizon, it remains the same for the entire ensemble generated by $\xi_A$ (note that on the horizon the vector fields $\xi_A$ are tangential to horizon). The choice of the two conditions defining the vector fields $\xi_A$ is guided by two obvious facts: one, the ensemble must be small enough so that the vector fields $\xi$ are sufficiently constrained and two, the ensemble must be large enough so as to include a large class of known black hole space-times. 

Since $A(\tau)$ is a smooth periodic function $A(\tau)=A(\tau+\beta)$ where $\beta=2\pi/\kappa$, it can be expanded in the Fourier modes $A_n(\tau)=\frac{1}{\kappa}\exp(in\kappa\tau)$ where $n$ is an integer (the normalizing factor $1/\kappa$ is chosen for a later convenience). This gives rise to an infinite number of vector fields $\xi_n\equiv\xi_{A_n}$, one associated with each Fourier mode, which obey the Witt algebra, $i[\xi_n,\xi_m]=(n-m)\xi_{n+m}$ (in general, $[\xi_A,\xi_B]=\xi_{A\del_\tau B-B\del_\tau A}$). The zero mode $\xi_0$ is essentially the Killing field $T$.

We now extend the vector field (\ref{xi}) from the $(R,T)$-plane to the full space-time neighborhood $\mfs N$ (we have already assumed that $\mfs N$ contains a Killing horizon of topology ${\bf R}\times{\bf S}^2$). At this point, it is necessary to make some further assumptions: {\tt (i)} $\xi_n$ remain tangential to the $(R,T)$-plane, {\tt (ii)} let $\mfs N$ possesses another Killing field $\Phi$ that generates axisymmetry (this invariably picks up another coordinate $\phi$ by $\Phi=\del/\del\phi$). The first assumption is well-motivated because we want to preserve the plane. However, the second assumption is more crucial because it restricts the ensemble to axially symmetric (at least in the neighborhood $\mfs N$) space-times only. Although this assumption is mild in the sense that a large number of interesting black hole space-times are axially symmetric, it appears to be unavoidable at this stage. We think that there is a way around this second assumption, namely it will 
be possible to include space-times having no symmetry at all other than the Killing symmetry $T$ only (in some neighborhood $\mfs N$ of the horizon), but we want to keep this development for some future work. Since no other coordinates (except $\tau,\phi$) can be canonically chosen in $\mfs N$, any extension of the vector field $\xi_n$ depending on coordinates other than $\tau,\phi$ would fail to be a canonical extension (namely such extensions will depend on the choice of these coordinates). To avoid this, let $\xi_n$ or more precisely the function $A$ depend only on the two coordinates $\tau,\phi$ which are canonically chosen by the local Killing vectors. Since $\kappa\tau$ and $\phi$ are two angles, there are two closed orbits generated by the diffeomorphisms: one, whose tangent vector field is $\lie_\xi T/\kappa$ and the other, who tangent vector field is $\lie_\xi\Phi$. It is natural that the two orbits are identical, {\rm i.e.}, $\lie_\xi T/\kappa=\lie_\xi\Phi$. Actually, by introducing another angle 
$\phi$ we are making a loop algebra extension of the Witt algebra. Since a loop extension is completely characterized by another integer, in general a pair of integers is involved -- an integer $n$ for the Witt algebra and another integer $m$ characterizing the loop extension and the new Fourier modes are $A_{nm}=\frac{1}{\kappa}\exp(in\kappa\tau+im\phi)$. Identifying the two orbits amounts to choosing $m=n$, a diagonal sub-algebra of the loop extended Witt algebra. This leads us to the following form for the vector fields
\beq \xi_n=A_n\frac{\partial}{\partial\tau}-\omega\del_\tau A_n\frac{\partial}{\partial\omega},\quad A_n=\frac{1}{\kappa}e^{in(\kappa\tau+\phi)}.\label{symmvec}\eeq
We claim that (\ref{symmvec}) is the right symmetry generating vector fields close the horizon.

Next, we construct the ensemble of space-times that remains invariant under the diffeomorphisms generated by these symmetry generating vector fields close to the horizon. Clearly, only the near-horizon form of the space-times can be restricted and the geometry of the space-times in other asymptotic regions will remain arbitrary and unspecified. Consider an ensemble of sufficient generality (motivated by the near horizon geometry of the Kerr space-time) such that it has the near horizon form
\begin{align} ds^{2}=\kappa^2\omega^2d\tau^2&+d\omega^2
+2\omega^{2}N(\theta)d\tau d\phi+2\omega^{2}\bar{N}(\theta)d\tau d\theta\nn
&+2\omega M(\theta)d\omega d\theta+2\omega\bar{M}(\theta)d\omega d\phi\nn
&+X(\theta)d\theta^2+2Y(\theta)d\theta d\phi+Z(\theta)d\phi^2\nn &+\text{higher order terms,}\label{ensemble}
\end{align}
where $N,\bar{N},M,\bar{M},X,Y,Z$ are arbitrary smooth functions and $\kappa$ is a constant fixed for the entire ensemble. For each choice of these functions, we get a class of space-times (because their behavior at other asymptotic regions may be different) that have identical near-horizon geometry. Each space-time admits a Killing horizon and the integral of the two form $\sqrt{XZ-Y^2}\,d\theta\wedge d\phi$ over the compact surface ${\bf S}^2$ of the horizon gives the area of the horizon. There is only one $o(1)$ structure at the horizon, namely the surface gravity $\kappa$ which is fixed in the class of space-times defined by (\ref{ensemble}). We are going to show now that this ensemble remains invariant under the diffeomorphisms generated by $\xi_A$. In fact, $\xi_A$ generates near-horizon symmetry in this precise sense. It may be worth looking for the most general ensemble of space-time that remains invariant under $\xi_A$, but for the present purpose we will work with this ensemble. The well known 
examples, such as Kerr and Schwarzchild space-times belong to this class (see appendix \ref{app:nhkerr} and \ref{app:nhschw}). The action of $\xi_A$ (\ref{symmvec}) on the ensemble generates the following fall-offs for the other metric components (we have already seen that $\xi_A$ preserves the $(R,T)$-plane, in fact this is how we got the vector fields $\xi_A$ in the first place):
\beq \delta_\xi g_{\tau a}=o(\omega^2),\quad \delta_\xi g_{\omega a}=o(\omega),\quad \delta_\xi g_{ab}=o(\omega^2).\eeq
where $a,b=\theta,\phi$. Clearly, the fall-offs exhibit just the right orders so that the diffeomorphisms generated by $\xi_A$ map the ensemble to itself.

To find a representation of this algebra in phase space we need to choose a theory.  We choose the simplest Einstein-Hilbert Lagrangian for gravity without matter. The Lagrangian induces a symplectic structure and the Poisson bracket of the associated Hamiltonians $\xi_n\mapsto H_n$ is given in \cite{silva} (Here, we are making a technical assumption which is verifiable, namely whether $\xi_n$s are Hamiltonian vector fields under the boundary conditions that the near-horizon geometry is restricted to (\ref{ensemble}). However, we proceed here assuming that this is the case and that the Poisson bracket of these Hamiltonians is given by the following expression.)
\begin{align} [H_n,&H_m]=\frac{1}{16\pi G}\int\varepsilon_{abcd}[\grad^c\xi_m^d\grad\cdot\xi_n-\grad^c\xi_n^d\grad\cdot\xi_m\nn
&+2\grad_e\xi^c_m\grad^e\xi^d_n+(R^{cd}{}_{ef}-2\delta^{c}_{e}R^{d}_{f})\xi^e_m\xi^f_n] \label{PoB}
\end{align}
where the integration is carried out over the two-sphere of the horizon. In order to evaluate this Poisson bracket explicitly we have to pick a candidate space-time $\mfs M$ (again, only the near-horizon geometry of $\mfs M$ will be relevant) from the ensemble (\ref{ensemble}). Putting (\ref{symmvec}) back into the expression (\ref{PoB}), a long but straightforward computation gives a central extension
\beq i[H_n,H_m]=\frac{n^3A}{8\pi G}\,\delta_{m+n}\label{result}\eeq
where
\beq A=\int \sqrt{XZ-Y^2}\,d\theta\wedge d\phi \eeq
is the area of the horizon. Interestingly, the result is independent of the functions $N,\bar{N},M$ and $\bar{M}$ and depends on the choice of $X,Y,Z$ only through the combination $X Z-Y^2$. This shows that any other space-time in the ensemble sharing the same horizon-area, the Poisson bracket will remain the same -- in other words, for each space-time $\mfs M$ belonging to the ensemble (\ref{ensemble}) there exists an area sub-ensemble (in the sense that each space-time in the area-ensemble has a fixed common horizon-area $A$) and the central extension is completely characterized by this sub-ensemble. The area ensemble is completely analogous to the standard fixed energy micro-canonical ensemble in statistical mechanics. We will see that it is this area-ensemble that provides an entropy to a candidate space-time $\mfs M$. So although Kerr or Schwarzschild space-times belong to the same big ensemble (\ref{ensemble}) the area sub-ensembles in which the two space-times belong are disjoint (For an 
argument see appendix \ref{app:nhkerrvsndschw}).

The Poisson bracket algebra (\ref{result}) is isomorphic to the Virasoro algebra if and only if the zero mode of the on-shell Hamiltonian is non-vanishing, namely $H_n=h\delta_{n,0}$. This has been found to be the case in all previous analysis \cite{Carlip1,OAW}. We assume that this holds for our Hamiltonian also. This is expected from fairly general arguments -- the zero mode is essentially the Killing vector field which is the only genuine symmetry of the background metric and so $H_0$ is some quasi-local mass of the space-time at the horizon; whereas the nonzero modes generate pure diffeomorphisms and the associated Hamiltonians generate gauge transformations, hence they should vanish on-shell (although their Poisson brackets are non-trivial). If this condition holds then it is possible to cancel the linear term $(n-m)H_{n+m}=2nh\delta_{n+m}$ of the Virasoro algebra with the central linear term $-(c/12)n\delta_{n+m}$ by choosing $h=c/24$. Now the Virasoro algebra contains only the cubic term $(c/12)n^3\delta_{n+m}$ which can be compared with the cubic term (\ref{result}). This gives $c=3A/(2\pi G)$ and hence also $h$. Another equivalent way to get the zero mode and the central charge is to calculate the simplest non-trivial Poisson brackets, first by assuming that the $H_n$s satisfy the Virasoro algebra and then using (\ref{result}). This gives
\begin{align} &i[H_1,H_{-1}]=2H_0=\frac{A}{8\pi G}, \\ &i[H_2,H_{-2}]=4H_0+\frac{c}{2}=\frac{A}{\pi G},\end{align}
which gives the same $H_0=A/16\pi G$ and $c=3A/2\pi G$. Then using the Cardy formula \cite{Cardy} we get
\beq S=2\pi\sqrt{\frac{cH_0}{6}}=\frac{A}{4G}.\eeq
The correct factor of $\hbar$ in the entropy is reproduced by the naive quantization $[H_n,H_m]=i\hbar[H_n,H_m]_{\rm PB}$. Since a pair $H_0$ and $c$ (here, given by the horizon-area $A$) completely characterizes an irreducible representation of the Virasoro algebra, it is the area-ensemble in which such a representation is furnished. Therefore, in the entropy, only the states associated with the space-times belonging to the area-ensemble are counted.

\section{Comparison}
At this point a detailed comparison between our analysis and the earlier ones (including ours in \cite{OAW}) is necessary. The main difference between the present analysis and the earlier ones is that our vector fields have smooth limits to the horizon; whereas to the best of our knowledge all the previous authors have considered such vector fields that do not have smooth limits to the horizon. Another major improvement is that we have clarified the role of the ensembles, a large one that is kept invariant under the action of these vector fields through diffeomorphisms and a smaller ensemble which we called the area-ensemble which arises in the context of a fixed background space-time and plays a central role in the calculation of entropy. Probably, it is the first time that in the method of the symmetry based approaches to black hole entropy the role of the two ensembles has been clarified. Our analysis shows the importance of the $(R,T)$ - plane.  The assumptions that we have made to single out the vector fields $\xi$ have been formulated solely in terms of this plane and the Killing vector fields of the space-time. In our derivation we have made use of the Killing vector $\partial_\phi$.  It is currently not clear if our analysis can be extended to space-times that do not possess such a symmetry. However, all earlier analysis seem to possess the same weakness. We hope that this is not a major obstacle and the analysis can be extended to cases that do not involve this additional symmetry. 



The fall-off conditions that Carlip has used possess some similarities with our fall-off conditions but his choices eventually led to vector fields that do not have smooth limits to the horizon. Moreover, it was shown that these original conditions were not enough for the algebra of the vector fields to close \cite{OAW}. However, even with the corrected conditions, the vector fields remain divergent at the horizon. 

It is to be noted that some authors, for instance Carlip, got an entropy $S=\sqrt 3A/4G$ from the Cardy formula. They then used a shift $H'_{0}=H_{0}-c/24$ (such a shift is permitted by the algebra) to get the entropy $S=\sqrt 2A/4G$ which still differs from the Bekenstein-Hawking entropy by a factor of $\sqrt 2$. This point was clarified in \cite{silva} where it was pointed out that using the corrected Poisson removes this problem. This is the reason why we used the Poisson bracket of \cite{silva}. However, it remains an open problem to verify that a symplectic structure exists in Einstein's theory in the presence of the boundary conditions imposed by (\ref{ensemble}) and that the symplectic structure leads to the Poisson bracket (\ref{PoB}). It also remains to be seen that the symmetry generating vector fields $\xi_n$ are Hamiltonian under the symplectic structure. These issues will be investigated in future work.

\section{Conclusion}

All Killing horizons share a common near-horizon feature -- the $(R,T)$-plane that captures the essence of a black hole. The Killing vector $T$ and the vector $R$ (that is constructed from the normal derivative of the square-norm of $T$) span this two dimensional plane. The remarkable fact is that for all Killing horizons this plane carries a universal metric, which is parametrised by the surface gravity $\kappa$ of the horizon only. Given a background metric, its $\kappa$ is fixed and there is no ambiguity in defining the area-ensemble. However, if the full background metric is not known and only its near-horizon form is given then the surface gravity $\kappa$ may not be unique and can differ by a $o(1)$ constant since $T$ is Killing only in some neighborhood of the horizon. In order to have a unique $\kappa$ in this case we have to fix the asymptotic behavior of the metric. This does not affect any of our results because the results are not sensitive to the behavior of (\ref{ensemble}) at asymptotic infinity. However, for the purpose of having a unique $\kappa$ for the ensemble we can take the space-times to be asymptotically flat so that at infinity $T$ can be identified with the unique time-translation vector field.  

Given that all Killing horizons share this $(R,T)$-plane and all horizons have an entropy it seems natural to ask whether one can find an explanation of black hole entropy solely in terms of the $(R,T)$-plane. This is exactly what we have investigated in this paper. We find a natural set of smooth vector fields in the plane which form an algebra that is isomorphic to the diffeomorphism algebra of the circle. When represented on the phase space of Einstein's theory, this algebra acquires a central extension that makes the diffeomorphisms generated by these vector fields a physical symmetry. If we assume that the black hole space-times form a representation for this physical symmetry, we get an entropy (using Cardy the formula) associated with these Killing horizons of some fixed area. This gives a universal explanation for black hole entropy.  

Black holes are often viewed as the Hydrogen atom for quantum gravity. The thinking goes as follows: if one is able to find an explanation for black hole entropy in terms of an underlying quantum theory that theory may be considered to be a candidate of quantum gravity. However, in the symmetry based approaches one is able to derive black hole entropy from purely semi-classical considerations. This leads us to doubt whether the question of black hole entropy plays any role in deciding whether a quantum theory of gravity is correct or not. If the theory allows for semi-classical states representing black holes it will also give the correct black hole entropy. Also, in any quantum theory of black holes the semi-classical states should belong to a dense subset of the full Hilbert space on which the symmetry algebra is faithfully represented.

Given that there are a number of candidate theories of quantum gravity, it would be interesting to see if the operators $H_n$ that we have found in the classical theory can be constructed as quantum mechanical operators $\hat H_n$ acting on the Hilbert space of a quantum black hole. Investigations in this direction are under way by the present authors.

\appendix

\section{Near-horizon Kerr metric}\label{app:nhkerr}
The metric of Kerr black hole in the Boyer-Lindquist coordinates is given by:

\beq ds^2=-\left(\frac{\Delta-a^2\sin^2\theta}{\Sigma}\right)d\tilde{t}^2\\
-\frac{2a\sin^2\theta(r^2+a^2-\Delta)}{\Sigma}d\tilde{t}d\tilde{\phi} \nonumber \\   +\left( \frac{(r^2+a^2)^2-\Delta a^2\sin^2\theta}{\Sigma}\right)\sin^2\theta d\tilde{\phi}^2 \nonumber \\
+\frac{\Sigma}{\Delta}dr^2+\Sigma d\theta^2 \nonumber \eeq
where
\beq\Delta=r^2+a^2-2Mr\eeq
\beq\Sigma=r^2+a^2\cos^2\theta\eeq
The two Killing horizons are located at
\beq r_{\pm}=M\pm\sqrt{M^2-a^2}\eeq
The killing vector corresponding to horizon at $r_{+}$ is given by:
\beq T^{a}=\partial_{\tilde{t}}^a+\Omega\partial_{\tilde{\phi}}^a\eeq
where the angular velocity $\Omega$ is:
\beq \Omega=\frac{a}{r_{+}^2+a^2}\eeq
and the surface gravity is given by:
\beq\kappa=(r_{+}-r_{-})\frac{\Omega}{2a}\eeq 
Calculating $T^2$
\beq -\omega^2=T^2=-\frac{\Delta}{\Sigma}(1-\Omega a\sin^2\theta)^2 \nonumber \\
+\frac{\sin^2\theta}{\Sigma}(r^2+a^2)^2\left(\frac{a}{r^2+a^2}-\Omega\right)^2\eeq
Setting $\Sigma_{+}=\Sigma(r_{+})$ 
\beq 1-a\Omega \sin^2\theta=\frac{\Sigma_{+}}{r_{+}^2+a^2}\eeq
we get
\beq -\omega^2=T^2=-\frac{\Delta}{\Sigma}\left(\frac{\Sigma_{+}}{r_{+}^2+a^2}\right)^2 \nonumber \\
+\frac{\sin^2\theta}{\Sigma}(r^2+a^2)^2\left(\frac{a}{r^2+a^2}-\Omega\right)^2\eeq
Neglecting higher order terms in $r-r_{+}$:
\beq \omega^2=\frac{2\kappa\Sigma_{+}}{r_{+}^2+a^2}(r-r_{+})\eeq
where expansion of $\Delta /\Sigma$ near horizon has been used:
\beq\frac{\Delta}{\Sigma}=\frac{2\kappa(r_{+}^2+a^2)}{\Sigma_{+}}(r-r_{+})\eeq
Following expression (44):
\beq 2\omega d\omega=\frac{2\kappa\Sigma_{+}}{r_{+}^2+a^2}dr\nonumber \\
-\frac{r-r_{+}}{r_{+}^2+a^2}2\kappa a^2\cos\theta \sin\theta d\theta\eeq
Now we introduce co-ordinates such that:
\beq \partial_{t}^a=T^a\eeq
This is achieved by the following transformations:
\beq t=\tilde{t}\eeq
\beq\phi=\tilde{\phi}-\Omega \tilde{t}\eeq
or
\beq\tilde{\phi}=\phi+\Omega t\eeq
writing the metric in co-ordinates $t,\phi,\omega$. $g_{tt}$ is:
\beq g_{\tilde{t}\tilde{t}}+2\Omega g_{\tilde{t}\tilde{\phi}}+\Omega^2g_{\tilde{\phi}\tilde{\phi}}=T^2=-\omega^2
\eeq
For the $d\omega$ term:
\beq \frac{\Sigma}{\Delta}dr^2=\frac{\Sigma_{+}}{2\kappa(r_{+}^2+a^2)(r-r_{+})}\nonumber
\\\times \left(\frac{r_{+}^2+a^2}{2\kappa\Sigma_{+}}2\omega d\omega-\frac{r-r_{+}}{2\kappa\Sigma_{+}}2a^2\cos\theta \sin\theta d\theta\right)^2\eeq
\beq=\left(\frac{d\omega^2}{\kappa^2}+2\omega M(\theta)d\omega d\theta+o(\omega^2)d\theta^2\right)\eeq
the $dtd\phi$ term:
\beq-\frac{2a\sin^2\theta(r^2+a^2-\Delta)}{\Sigma}dtd\phi+\nonumber \\
\left( \frac{(r^2+a^2)^2-\Delta a^2\sin^2\theta}{\Sigma}\right)\sin^2\theta\ 2\omega dtd\phi \eeq
Now at $r=r_{+}$ this term is zero, so to lowest order this term can be $o(\omega^2)$. So this term can be written as $\omega^2 X(\theta)$. $g_{\theta\theta}$ is already $o(1)$, so the contribution from (53) can be neglected. So, this term can be written as $Y(\theta)$.The term $g_{\phi\phi}$ is $o(1)$, so can be written as $Z(\theta)$. Therefore the metric (23) follows.

\section{Near-horizon Schwarzschild geometry}\label{app:nhschw}
The metric for near-horizon Schwarzschild is:
\beq ds^{2}=\omega^2d\tau^2+\frac{d\omega^2}{\kappa^{2}}
+4M^2(d\theta^2+\sin^2\theta d\phi^2)
\eeq

\section{Kerr and Schwarzschild sub-ensembles}\label{app:nhkerrvsndschw}

The metric on the two-sphere for Schwarschild is:
\beq ds^2=4M^2(d\theta^2+\sin^2\theta d\phi^2) \eeq 
which admits three killing vectors. Now since the Killing equations are invariant under any diffeomorphism, if a background metric possesses a Killing vector then by the action of diffeomorphisms one can not generate a space-time that possesses a different isometry. Since the area-ensemble in which the Virasoro algebra is expected to furnish an irreducible representation, any two space-times of the area-ensemble are related by some diffeomorphism, i.e. the Virasoro generators $H_n$ act transitively on the area-ensemble. This shows that the area-ensemble for the Schwarzschild and Kerr space-times are necessarily disjoint. Similarly, Kerr and extremal Kerr are expected to belong to two different ensembles (in this paper we have not analysed the extremal $\kappa=0$ space-times on the basis of symmetry based approaches, however, we hope to analyse this case in some future work).      





\begin{acknowledgments}
Olaf Dreyer would like to thank the Foundational Questions Institute (FQXi) for partial support during this work. Amit Ghosh would like to thank the Max Planck Institute for Gravitational Physics (AEI), Potsdam, Germany, for its kind hospitality where part of the work was done and Daniele Oriti, Daniele Pranzetti, Isabeau Premont-Schwarz, Jose-Luis Jaramillo and Mohab Abou Zeid of AEI for discussion.  
\end{acknowledgments}


\end{document}